\documentclass[review=false,timestamp=false,nonacm,screen]{acmart}
\AtBeginDocument{%
  \providecommand\BibTeX{{%
    \normalfont B\kern-0.5em{\scshape i\kern-0.25em b}\kern-0.8em\TeX}}}

\begin{document}
\title[Open MatSci ML Toolkit]{Towards Foundation Models for Materials Science: The Open MatSci ML Toolkit}

\author{Kin Long Kelvin Lee}
\email{kin.long.kelvin.lee@intel.com}
\orcid{0000-0002-1903-9242}
\affiliation{%
  \institution{Accelerated Graphics \& Computing Systems, Intel Corporation}
  \streetaddress{2111 25th NE Ave.}
  \city{Hillsboro}
  \state{OR}
  \country{USA}
  \postcode{97124}
}
\author{Carmelo Gonzales}
\email{carmelo.gonzales@intel.com}
\affiliation{%
  \institution{Intel Labs, Intel Corporation}
  \streetaddress{2200 Mission College Blvd.}
  \city{Santa Clara}
  \state{CA}
  \country{USA}
  \postcode{95054}
}
\author{Matthew Spellings}
\email{mspells@vectorinstitute.ai}
\affiliation{%
  \institution{Vector Institute}
  \city{Toronto}
  \state{ON}
  \country{Canada}
}
\author{Mikhail Galkin}
\email{mikhail.galkin@intel.com}
\affiliation{%
  \institution{Intel Labs, Intel Corporation}
  \streetaddress{2200 Mission College Blvd.}
  \city{Santa Clara}
  \state{CA}
  \country{USA}
  \postcode{95054}
}
\author{Santiago Miret}
\email{santiago.miret@intel.com}
\affiliation{%
  \institution{Intel Labs, Intel Corporation}
  \streetaddress{2200 Mission College Blvd.}
  \city{Santa Clara}
  \state{CA}
  \country{USA}
  \postcode{95054}
}
\author{Nalini Kumar}
\email{nalini.kumar@intel.com}
\affiliation{%
  \institution{Accelerated Graphics \& Computing Systems, Intel Corporation}
  \streetaddress{2200 Mission College Blvd.}
  \city{Santa Clara}
  \state{CA}
  \country{USA}
  \postcode{95054}
}

\renewcommand{\shortauthors}{Lee et al.}

\begin{abstract}

Artificial intelligence and machine learning have shown great promise in their ability to accelerate novel materials discovery. As researchers and domain scientists seek to unify and consolidate chemical knowledge, the case for models with potential to generalize across different tasks within materials science---so-called ``foundation models''---grows with ambitions. This manuscript reviews our recent progress with development of Open MatSci ML Toolkit, and details experiments that lay the groundwork for foundation model research and development with our framework. First, we describe and characterize a new pretraining task that uses synthetic data generated from symmetry operations, and reveal complex training dynamics at large scales. Using the pretrained model, we discuss a number of use cases relevant to foundation model development: semantic architecture of datasets, and fine-tuning for property prediction and classification. Our key results show that for simple applications, pretraining appears to provide worse modeling performance than training models from random initialization. However, for more complex instances, such as when a model is required to learn across multiple datasets and types of targets simultaneously, the inductive bias from pretraining provides significantly better performance. This insight will hopefully inform subsequent efforts into creating foundation models for materials science applications.

\end{abstract}






\maketitle

\section{Introduction}

Materials science---despite its relative youth as a scientific discipline---has become a dominant driver for modern day technology, with applications ranging from semiconductors in electronics, to battery materials and light harvesting in energy, to catalysts for faster, safer, and more efficient chemical manufacturing. Key to all of this is the discovery of new materials: as each of the aforementioned areas grow, the materials we use need to grow and adapt to meet demand. Until quite recently, the design of materials has relied exclusively on a feedback loop between experiment and theory: the former attempts to synthesis and characterize novel materials with desirable properties, while the latter seeks to understand underlying electronic structure to guide design.

The shortcomings of this conventional loop, however, generally pertain to scalability and cost. Experimental facilities are costly to start-up and run, and are typically limited in throughput: chemical synthesis procedures can span weeks and months, and the expertise to realize these procedures can be hard to cultivate. For this reason, computational modeling has traditionally complemented experimental work by enabling higher throughput screening, whereby properties of promising materials can be calculated either from first-principles (\textit{ab initio} methods such as coupled-cluster theory \cite{coesterShortrangeCorrelationsNuclear1960,matthewsCoupledclusterTechniquesComputational2020}) and/or semi-empirical/approximate methods such as density functional theory (DFT) \cite{kohnSelfConsistentEquationsIncluding1965}. While in some cases, computer hardware and algorithmic improvements \cite{nakataLargeScaleLinear2020} have rendered computation faster than experiment, the design of space of chemistry is so vast \cite{reymondChemicalSpaceProject2015} and the scaling of compute requirements is so costly that computational modeling is still reserved only for the most promising materials.

The advent of machine learning (ML) and artificial intelligence (AI) provides an avenue for significantly changing---hopefully improving---the way materials can be and are discovered. Advancements in machine-learned representations of atoms, learning algorithms (in particular neural network architectures and optimization), and high-performance computing jointly enable new routes to approach materials discovery. For example, various groups \cite{gasteigerGemNetUniversalDirectional2021,chenGraphNetworksUniversal2019,spellingsGeometricAlgebraAttention2022} have seen success in parameterizing deep neural networks as surrogate models for property prediction, which once trained, can be used for prediction at a fraction of the computational cost of conventional electronic structure methods. 

AI/ML workflows have their own requirements that need to be realized: high-quality and diverse datasets are needed to train \emph{useful} models, and new algorithms and architectures central to materials science should be developed. To do so requires uniting high performance computing with intuitive abstractions for materials scientists, chemists, and AI/ML practitioners and engineers to cooperate seamlessly. These are the guiding principles behind the development of Open MatSci ML Toolkit---a framework that seeks to unite large-scale and disparate materials datasets, state-of-the-art model implementations, and an end-to-end, modular pipeline that scales with use cases from developing and  testing locally on laptops to many-node distributed computing. An outcome of this is to facilitate general purpose AI/ML models for materials science---so-called ``foundation models'' \cite{bommasaniOpportunitiesRisksFoundation2022}---trained on large amounts of diverse data to produce unified representations \cite{rebuffiLearningMultipleVisual2017} that can be tuned to specific tasks with less data. 

The present paper provides a brief introduction to these concepts, and reviews some of the recent discoveries and challenges we have found towards this goal. Our key contributions include:

\begin{enumerate}
    \item A new pretraining task based on synthetic point clouds based off sets of symmetry operations,
    \item Demonstration of a distributed development, training, and experimentation workflow with up to 32 nodes of 4th Generation Intel\textregistered~Xeon\textregistered~Scalable Processors (codenamed Sapphire Rapids),
    \item Characterization of complex training dynamics during pretraining, which we ascribe to recent investigations into instabilities of the Adam optimizer,
    \item Investigation of qualitative and quantitative benefits of pretraining on downstream tasks, including under a multi-task, multi-dataset setting.
\end{enumerate}

\section{Background and related work}

\subsection{Atomistic representations}

There have been a host of methods for representing chemical structures in machine-readable formats, which are subsequently ingested by some AI/ML algorithm to transform the data into vector representations. Graph structures have been the workhorse of chemists for many years \cite{cayleyLVIIMathematicalTheory1874}, with atoms represented as nodes/vertices and connectivity between atoms can be arbitrarily structured as edges. The development of neural networks for graphs has grown rapidly, with chemical applications being at the forefront: graph neural networks (GNN) have frequently used molecular data \cite{ramakrishnanQuantumChemistryStructures2014} as proving grounds for learning symmetry-adapted properties that are equivariant \cite{satorrasEquivariantGraphNeural2022} or invariant \cite{schuttSchNetContinuousfilterConvolutional2017} to 3D geometry (i.e. translation, rotation, reflection).

A recent line of research extends these desirable properties to point clouds with geometric algebra \cite{spellingsGeometricAlgebraAttention2022, brehmerGeometricAlgebraTransformers2023}, which work in the space of multivector products. Some advantages of this approach over graphs is bypassing the need to impose structure, i.e. construction of the graph, on the data, which in turn has downstream implications such as computational performance; graph structures can exhibit poor cache reuse without reordering \cite{gongGraphiteOptimizingGraph2022}, in addition to the need for specialized sparse kernels \cite{heineckeLIBXSMMAcceleratingSmall2016}. In contrast, geometric algebra networks rely on generally optimized dense compute in combination with the well-studied attention mechanism, provides comparable modeling capabilities for small graphs.

\subsection{Related Work}

\subsubsection{Open Catalyst Project}

The Open Catalyst Project (OCP) \cite{chanussotOpenCatalyst20202021,tranOpenCatalyst20222022} was developed in collaboration by researchers at Fundamental AI Research (FAIR) at Meta AI and Carnegie Mellon University's Department of Chemical Engineering. It comprises a set of benchmark datasets and challenges associated with catalyst activity, whereby a molecular adsorbate interacts with a catalytic surface. At a high level, the tasks and datasets were constructed to parameterize graph neural networks as learned potentials for catalyst simulation. OCP has since motivated a number of related projects including state-of-the-art models \cite{gasteigerGemNetUniversalDirectional2021}, data-centric transformations \cite{duvalPhASTPhysicsAwareScalable2023}, and indeed, our own project Open MatSci ML Toolkit. Open MatSci ML Toolkit differentiates from OCP by expanding the scope in data representations (i.e. use of point clouds), tasks (property prediction and classification), and datasets.

\subsubsection{MatBench}

MatBench \cite{dunnBenchmarkingMaterialsProperty2020} provides a suite of evaluation benchmarks using structures and targets provided by the related Materials Project \cite{jainCommentaryMaterialsProject2013}, with various tasks including property prediction and material classification. Similar to OCP, communities are able to submit to leaderboard run by the authors, allowing new architectures and strategies to be demonstrated; to date, most submissions have involved GNN-based architectures. Open MatSci ML Toolkit uses the same data source (Materials Project) as MatBench, but differs in approach by allowing free composition of tasks and datasets as opposed to independent benchmarks, allowing knowledge to be pooled across areas in materials science.

\subsubsection{MAST-ML}

MAST-ML \cite{jacobsMaterialsSimulationToolkit2020} provides a platform for materials science research mainly with classical ML techniques in familiar notebook environments. The advantages of this approach is a low barrier to entry for practitioners, as well as rapid reproducible research and development by providing interfaces to several community data repositories such as the Materials Data Foundry \cite{blaiszikMaterialsDataFacility2016,blaiszikDataEcosystemSupport2019} and abstractions for scikit-learn models \cite{scikit-learn}. Open MatSci ML Toolkit differs by focusing mainly on deep learning techniques, which generally possess a higher degree of expressivity and capacity to learn than classical ML models like gradient boosting \cite{friedmanGreedyFunctionApproximation2001}. Perhaps the most striking similarity between the two projects is the ability to interact with different datasets, which contrasts with the other two related benchmarks.

\section{Open MatSci ML Toolkit}

As we have mentioned in the introduction, the Open MatSci ML Toolkit was designed to be an open source platform for AI-accelerated materials discovery, applying concepts ranging from MLOps best practices to supporting various representations of material structures, as described in the previous section. We direct interested readers to an earlier introduction to the toolkit \cite{miret2023the} and for brevity we will focus on some newer and relevant aspects to results we will discuss in this paper. The original concept for Open MatSci ML Toolkit was to improve performance portability of the Open Catalyst Project by refactoring significant amounts of functionality to PyTorch Lightning \cite{Falcon_PyTorch_Lightning_2019} and the Deep Graph Library (DGL): the former provides pipeline abstractions that facilitate accelerator offloading, scale-up and scale-out, and the latter offers GPU \emph{and} CPU-optimized sparse kernels for efficient graph neural network compute \cite{wang2019dgl}. Since the original release, we have significantly reworked the scope of the framework to move beyond Open Catalyst tasks and datasets, and provide a generalized abstraction to incorporate and maintain interfaces to a variety of data sources and modeling tasks.

\begin{figure}
    \centering
    \includegraphics[width=0.75\linewidth]{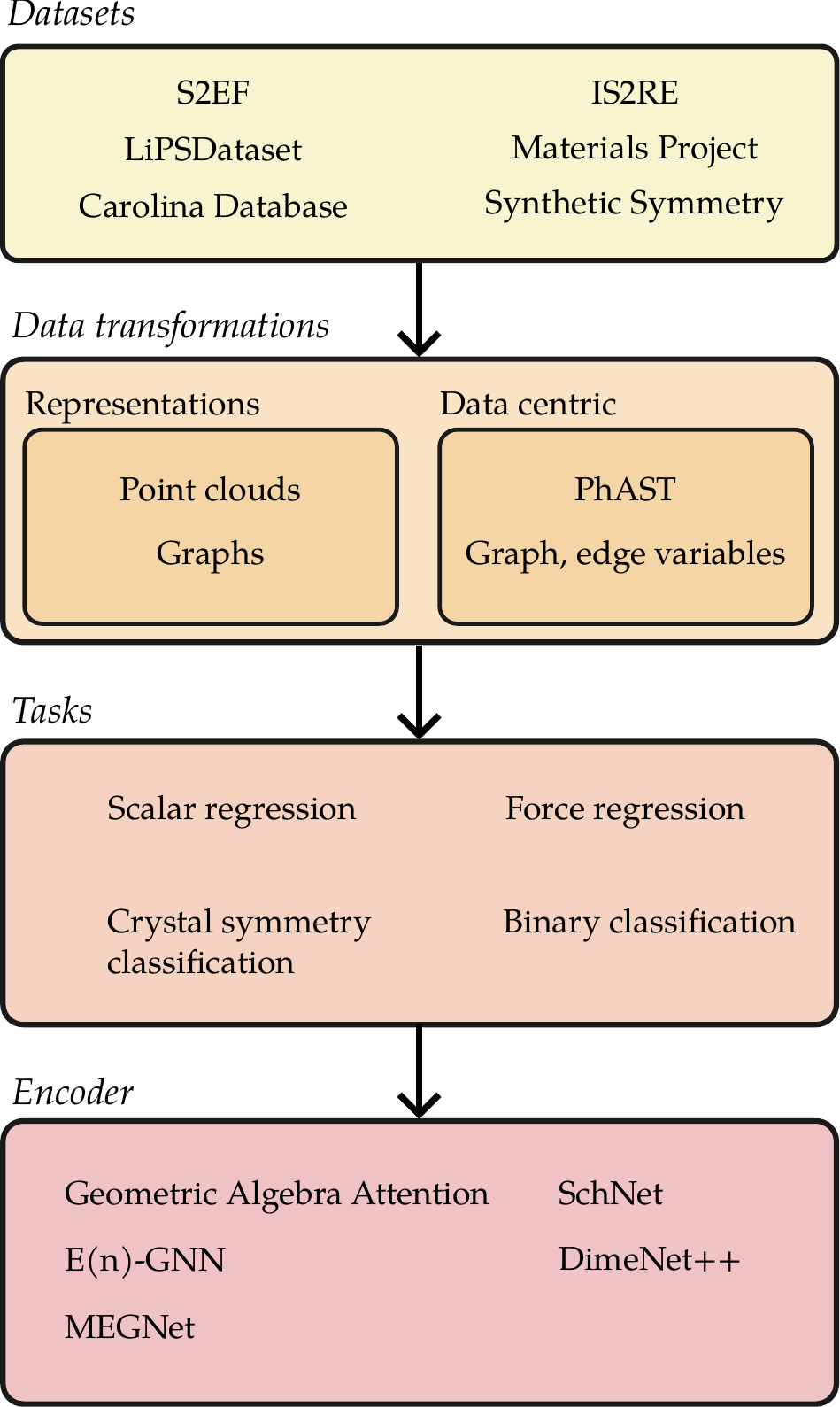}
    \caption{Modular workflow and abstraction for Open MatSci ML Toolkit, with some examples provided in each block. A number of widely used materials science \emph{datasets} have been incorporated, providing the foundations for experimentation. As each data sample is retrieved, a chain of \emph{transformations} can be applied to freely convert between representations, and/or modified to introduce inductive biases. A \emph{task} adapts the emitted data for a particular set of learning targets, and comprises an \emph{encoder} that processes the material to generate a vector embedding that is subsequently used by individual output heads for task predictions.}
    \Description{A diagram depicting how Open MatSci ML Toolkit was designed and can be used.}
    \label{fig:flowchart}
\end{figure}

\subsection{Datasets}

As of writing, Open MatSci ML Toolkit supports datasets that can be used for both property prediction and pretraining. For the former, we provide interfaces for a diverse set of datasets, including: the Open Catalyst Project \cite{chanussotOpenCatalyst20202021,tranOpenCatalyst20222022}; the Materials Project \cite{jainCommentaryMaterialsProject2013}; the Carolina Materials Database \cite{zhaoHighThroughputDiscoveryNovel2021, danGenerativeAdversarialNetworks2020}; and the LiPS dataset \cite{batznerEquivariantGraphNeural2022a}. These datasets range from material properties such as band gap and formation energy, as well as time-dependent dynamics with energy/force labels for trajectory samples. 

For pretraining, we include a pipeline for generating synthetic point clouds based off symmetry groups relevant to crystal structures: at a high-level, this involves creating a collection of particles in 3D space, followed by repetition based on randomly chosen symmetry operations. While these data do not necessarily reflect real material structures nor contain chemical information, they share the same symmetry basis, and by learning to identify point groups, we believe should instill some structural concepts in embeddings. Further, as the positions of particles are randomly generated, we are able to sample configurations at arbitrary scales and uniformly across point groups; this is especially important as materials datasets typically focus on particular types of structures and symmetries, thereby inducing a selection bias.

From Figure \ref{fig:flowchart}, the transformation interface and pipeline allows datasets to be freely converted between representations, as well as include additional variables and manipulations as needed by downstream modeling.

\subsection{Tasks}

Given a dataset and some neural network model, we define a ``task'' as a learning objective such as property regression (e.g. the semiconductor band gap), or material classification (e.g. is the material stable at 298\,K). Concretely, a task comprises an encoding model which ingests a graph or point cloud structure and emits an embedding that is subsequently used by one or several output heads (see Appendix \ref{sec:models}) to independently predict targets, in which the encoder and each output head is trained jointly through an optimizer. Each task is implemented as Python classes and can be freely paired with datasets in a modular way (Fig. \ref{fig:flowchart}); for instance in this work, the Materials Project dataset is used for both regression and classification tasks.

For more complex learning required for foundation models, Open MatSci ML Toolkit is able to seemlessly compose multiple datasets and tasks together, using an encoder model that learns generalized representations across them. On the one hand, multi-task learning makes full use of all labels present in a dataset, for example performing regression and classification jointly. On the other hand, using multiple datasets to perform a similar task creates the opportunity for models to specialize by scaling up available training data. Combining both multiple datasets and tasks in theory facilitates highly general models, providing the underlying encoder has the capacity and expressivity. In this case, the joint encoder is updated separately to each task output head.

\section{Experimental setup}

\subsection{Platform and software}

All experiments were carried out on the Endeavour cluster at Intel. Each compute node consists of two CPU sockets---both populated with Intel\textregistered~Xeon\textregistered~Platinum 8480+ processors comprising 56 physical cores partitioned into four non-uniform memory access (NUMA) nodes with combined 256\,GB of DDR5 (4800\,MT/s) memory.\footnote{Using Rocky Linux 8.8, kernel version \texttt{4.18}} Each compute node is connected via Mellanox HDR200 interconnect, and datasets stored on a Lustre filesystem. Scale-out experiments for measured throughput were conducted on 8/2/2023.

We used an internal build of Open MatSci ML Toolkit which we are preparing for open source release. The main dependencies include DGL (\verb|0.9.1|), PyTorch (\verb|1.13.1|), and PyTorch Lightning (\verb|1.8.6|). PyTorch was obtained as part of the Intel\textregistered~AI Kit \verb|2023.1| via the \verb|intel| \verb|conda| channel, which is prebuilt against optimized Intel\textregistered~oneAPI libraries including oneMKL and oneDNN, along with framework optimized libraries such as oneCCL Bindings for PyTorch. For intraop parallelism, we used Intel\textregistered~OpenMP threading, and for distributed training and inference, Intel\textregistered~MPI. In all instances of training and inference, processes were bound to their respective NUMA domains (\verb|-map-by numa|) with threads pinned to physical cores (\verb|I_MPI_PIN_CELL=core|).

\subsection{Training strategy}

To focus the scope of work, we only consider a single GNN architecture, the equivariant graph neural network [E(n)-GNN] developed by \citet{satorrasEquivariantGraphNeural2022}, with a nominal, untuned set of hyperparameters. Details regarding neural network architectures can be found in Appendix \ref{sec:models}.

In all cases, we use distributed data parallelism (DDP) to accelerate training: the full dataset is divided across MPI processes ($N$) receiving the same number of samples ($B$) per batch to yield an effective batch size $B_\mathrm{eff} = N B$. For a single node, the number of parallel workers---16 for our machine configuration---was chosen to balance compute per worker (i.e. the available FLOP/s determined by \verb|OMP_NUM_THREADS|) and memory bandwidth per socket, and interprocess communication only occurs at the gradient averaging step. In this work, we report scaling up to 32 nodes, yielding $N = 512$ and $B_\mathrm{eff} = 16,384$.

For all training, we use the Adam optimizer with decoupled weight decay (\verb|AdamW|) \cite{loshchilovDecoupledWeightDecay2019}, using default momentum values ($\beta_1=0.9$, $\beta_2=0.999$). For DDP training, the base learning rate ($\eta_\mathrm{base}$) is scaled by the total number of DDP workers to ensure constant gradient variance \cite{goyalAccurateLargeMinibatch2018}. Over the course of training, the learning rate $\eta$ is modified based on learning rate schedulers: a warmup period of eight epochs linearly ramps $\eta$ to its nominal value, followed by an exponential decay with $\gamma=0.8$. For fine-tuning, we scale $\eta_\mathrm{base}$ down by a factor of ten, which is typical practice to mitigate forgetting \cite{liLearningForgetting2017}.

\section{Results \& discussion}

\subsection{Training scale-out}

\begin{figure}
    \centering
    \includegraphics[width=\linewidth]{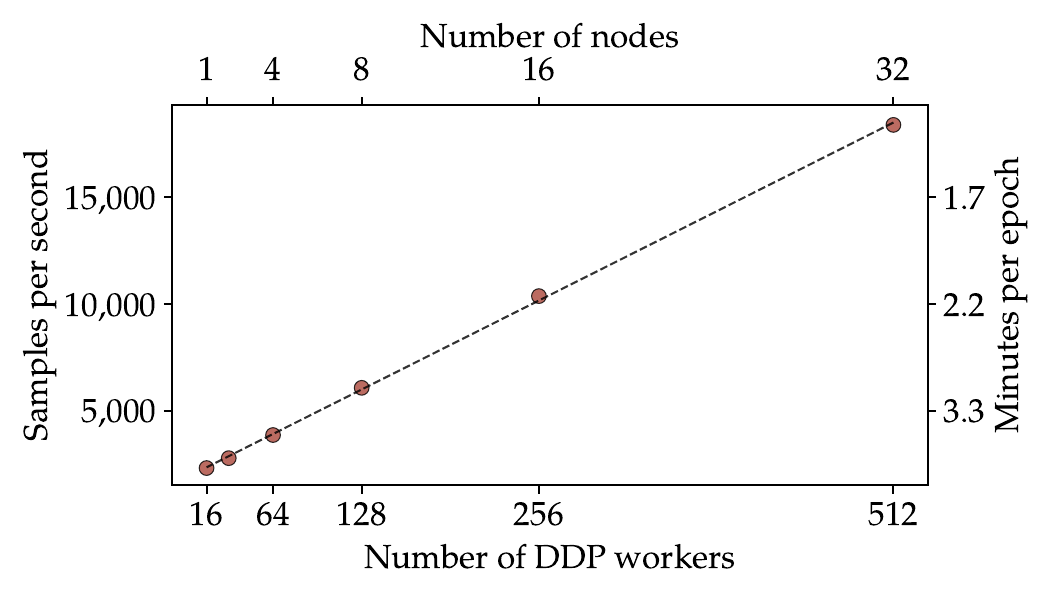}
    \caption{Pretraining throughput scaling as a function of the number of DDP workers, shown as both samples per second, and based on a dataset size of 2,000,000 samples, the time to complete an epoch on the right ordinate. The dashed line represents a linear fit to the measurements.}
    \label{fig:throughput-scaling}
\end{figure}

To facilitate high-throughput experimentation, we relied on both scaling-up and scaling-out training and inference to many MPI processes. Figure \ref{fig:throughput-scaling} shows the combined training throughput for the symmetry pretraining task as a function of DDP workers, as we scale out from one to a total of 32 nodes. We see a  linear growth in training throughput, implying that the communication overhead associated with gradient synchronization for every training step is negligible compared to the compute of individual workers. For additional context, the right ordinate of Figure \ref{fig:throughput-scaling} converts each tick for throughput into time per epoch; in most configurations considered, the full dataset is traversed within minutes, enabling quick feedback for experimental configurations.

\subsection{Symmetry group pretraining \label{sec:pretraining}}

To ensure quality of down stream modeling, it is important to characterize the convergence dynamics of pretraining, as well as the time to solution in a wall-time basis: higher training throughput can be achieved by scaling out to a larger number of DDP workers, however as the gradients are averaged over large $B_\mathrm{eff}$, fewer optimizer updates are performed with less information content. To partially compensate, we follow the recommendation of \citet{goyalAccurateLargeMinibatch2018} to scale the learning rate for a single batch proportional to the number of DDP workers (i.e. the world size).

\begin{figure}
    \centering
    \includegraphics[width=\linewidth]{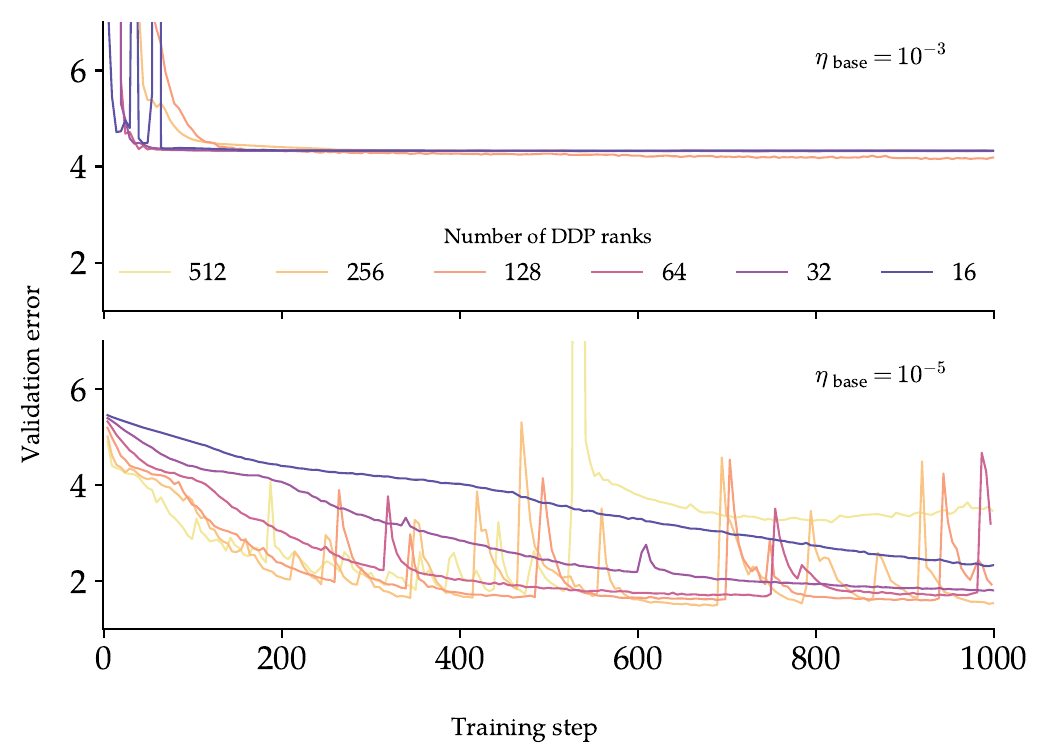}
    \caption{Early training dynamics for the pretraining task as a function of the number of DDP workers, which increases the effective batch size $B_\mathrm{eff}$ proportionally. The top and bottom frames correspond to higher and lower base learning rates. In both instances, the ordinate---representing the cross-entropy error---has been truncated to show scales relevant to the discussion.}
    \label{fig:training-dynamics}
\end{figure}

Figure \ref{fig:training-dynamics} shows the validation error as a function of the number of DDP workers for a fixed number of training steps. The top plot shows that, with $\eta_\mathrm{base}=10^{-3}$---a typical value for the learning rate---learning stagnates early for all scale-out configurations at large validation errors, indicating a combination of fairly broad local minima, and insufficiently informative gradients to escape them. An intuitive solution is to substantially lower the base learning rate, shown in the bottom plot of Figure \ref{fig:training-dynamics}. Here, while we see the validation error does decrease over time, we draw two general observations: first, the single node (16 DDP ranks) converges albeit slowly; second, the early rate of convergence seems to increase with the number of workers, however the prevalence of spikes in the validation loss also increases. In the largest scale-out case (512 ranks), the model experiences a large spike around step ${\sim}$550 and does not recover to former levels.

Recently, \citet{molybogTheoryAdamInstability2023} highlighted previously reported instabilities in the training dynamics of large language models \cite{chowdheryPaLMScalingLanguage2022}, and provided a thorough investigation into the convergence properties of Adam-based optimizers. The authors observe similar optimizer divergence, correlated with large gradient norms and attribute these occurrences to inherent instabilities in the Adam algorithm. In particular, one of assumptions required for convergent properties of Adam is zero time-correlation between update steps, i.e. Markovian dynamics, which are violated for large batch training, and result in a disconnect in the learning dynamics between layers whereby gradients decay to the order of $\varepsilon$ used to prevent division by zero. While \citet{molybogTheoryAdamInstability2023} discuss this in the context of large language models, the instabilities are general to ``Adam-like'' optimizers and likely to apply here, particularly as the use of GNN encoders can lead to over-smoothing (see \cite{ruschSurveyOversmoothingGraph2023} for a review) and therefore decrease signal contained within gradients; a symptom described by \citet{molybogTheoryAdamInstability2023} in early layers of language models. 

A thorough analysis into this phenomenon is outside our scope, and in the interim, we found a balance between convergence characteristics and training throughput with $N=256$, or 16 nodes. While this kind of optimization is conventionally offloaded to hyperparameter optimization, further work is required to assess a more principled approach for large scale pretraining in Open MatSci ML Toolkit. Subsequent experiments involve using a model trained for 20 epochs, taking approximately ${\sim}80$\,min.---an adequate turnover rate for experimentation.
% Unfortunately, training curves for related work have not been reported (for example, \cite{freyScalableGeometricDeep2021,freyNeuralScalingDeep2022a,sriramTrainingBillionParameter2022a}).

% From these observations, the symmetry pretraining task appears to demonstrate complexities in learning dynamics, which could be attributed to expressitivity in the neural network architecture or in the data itself. Given, however, this type of architecture has seen success in chemical tasks \cite{satorrasEquivariantGraphNeural2022} these explanations are unlikely. 

% \begin{enumerate}
%     \item Show training curves for the symmetry learning task
%     \item Show classification error for materials project structures without re-training
%     \item Show training curves of a property prediction task, with and without symmetry pretraining
% \end{enumerate}

\subsection{Dataset exploration with pretrained models}

Despite ongoing challenges with training, in this section we show that the model with learned representations of symmetry groups still possesses valuable inductive bias. One application of pretraining is the potential to architect datasets semantically based on structure, chemical composition, or both. Both elements reflect in the material design process, and thus the corresponding datasets used to train foundation models for materials science should include this aspect for composition. Here, we use the pretrained E(n)-GNN model from the previous section to embed samples contained across the datasets supported in Open MatSci ML Toolkit, and using UMAP \cite{mcinnesUMAPUniformManifold2018} as a structure-preserving projection method, visualize the distribution and diversity of material structures holistically. 

\begin{figure}
    \centering
    \includegraphics[width=\linewidth]{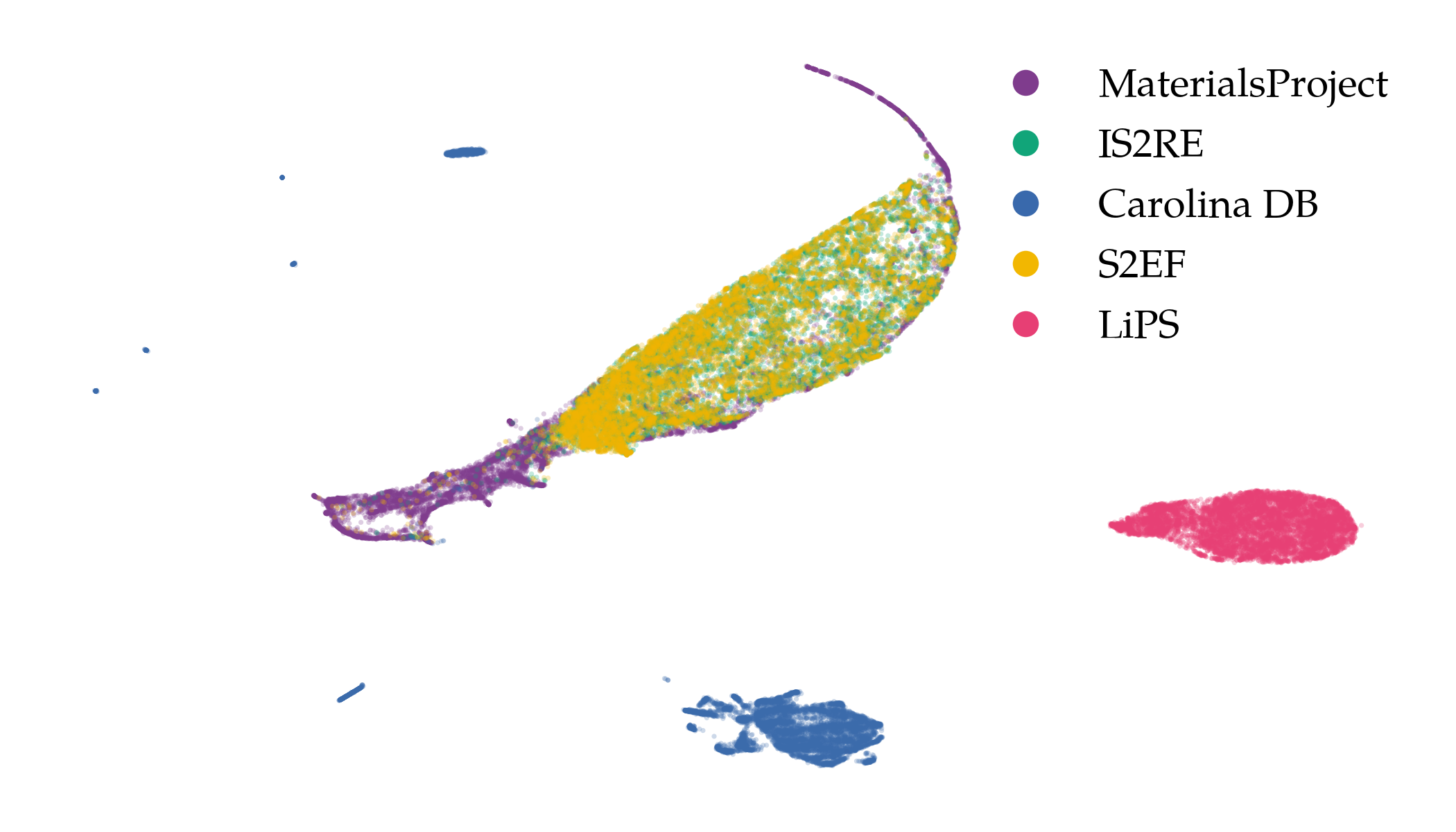}
    \caption{UMAP visualization of each dataset implemented in Open MatSci ML Toolkit. Each point represents a structure contained within its dataset, embedded with an E(n)-GNN encoder pretrained on synthetic constructs. For each dataset, 10,000 structures were randomly chosen as a representative sample. The UMAP parameters used were: 200 \texttt{n\_neighbors}, 0.05 \texttt{min\_dist}, using euclidean distance as the metric.}
    \Description{UMAP visualization of datasets implemented in Open MatSci ML Toolkit---this illustrates the space of structures contained.}
    \label{fig:structure-umap}
\end{figure}

The result is shown in Figure \ref{fig:structure-umap}, from which we can draw three qualitative observations: first, datasets tend to comprise the same kinds of structural motifs, which may be a reflection of inherent selection bias when it comes to composing datasets; second, somewhat expectedly, there is significant overlap between the OCP datasets; third, Materials Project data share some structural motifs with OCP, but perhaps offers the broadest variety of structures. For context, these observation can be calibrated by the fact that the LiPS dataset, which comprises trajectories of the same material composition, forms a clear independent cluster. The conclusions we can draw from these observations align well with our design philosophy: the inclusion of a variety of datasets provides the diversity of materials needed to complement one another, and this holistic view can provide valuable insight when considering what data to include, and how much is needed. The same analysis could be done using an encoder trained with chemical information, for example Materials Project, to find dataset gaps in \emph{chemical space}.

\subsection{Fine-tuning on material properties}

\begin{table*}[ht]
    \centering
    \caption{Comparison of task metrics for multi-task, multi-data training starting from a pretrained model, and from random initialization. With the exception of stability, which corresponds to the binary cross-entropy error, regression errors are provided as mean-absolute error in their respective units. Values in boldface correspond to the lowest error. $\zeta$ corresponds to the Fermi energy.}
    \begin{tabular}{l | c c c c | c}
        \toprule
        ~ & \multicolumn{4}{c|}{Materials Project} & Carolina Materials Database \\
        Configuration & Band gap (eV) & $\zeta$ (eV) & $E_\mathrm{form}$ (eV/atom) & Stability classification & $E_\mathrm{form}$ (eV/atom) \\
        \midrule
         Pretrained & \textbf{1.27} & \textbf{0.76} & \textbf{0.83} & 0.42 & 0.14 \\
         From scratch & 4.80 & 3.86 & 3.54 & \textbf{0.40} & \textbf{0.10} \\
         \bottomrule
    \end{tabular}
    \label{tab:wholehog}
\end{table*}

With a pretrained model, this section assesses whether there are quantiative benefits to pretraining for tasks relevant to materials discovery. We compare the validation performance of models that have been pretrained, versus from random initialization or ``from scratch'' for a given task. 

\begin{figure}
    \centering
    \includegraphics[width=\linewidth]{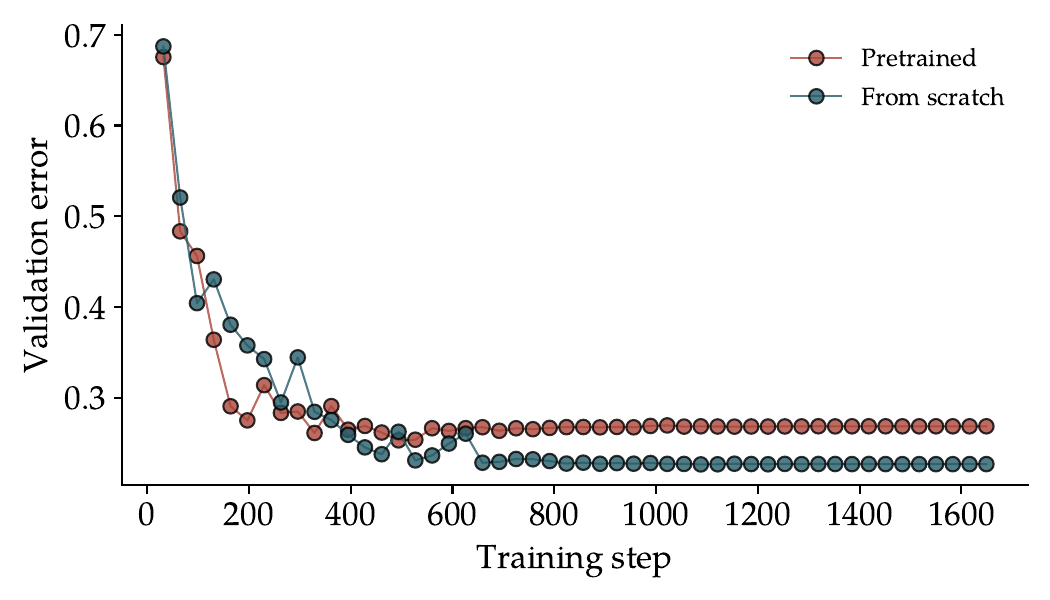}
    \caption{Validation curves for E(n)-GNN models trained to perform band gap prediction with the Materials Project dataset, comparing trajectories starting from a model pretrained on synthetic data (red) and from random initialization (gray blue).}
    \label{fig:bandgap}
\end{figure}

The simplest case, shown in Figure \ref{fig:bandgap}, corresponds to a band gap regression with the Materials Project Dataset. For both initializations, we observe smooth, convergent behavior. In the case of the pretrained model, we see that the early training stages converge to a lower error more quickly than the randomly initialized model, however appears to fall into a local minimum. In contrast, the model trained from scratch sees slower convergence, but eventually to a better performing model by the end of training. From this result, it appears that the pretraining may introduce a strong bias that allows for faster convergence that may see benefits with early stopping algorithms with a fixed compute budget, but for longer regimes, training from random initialization may provide sufficient momentum to escape local minima.

In a more complex procedure, and one that resembles more of foundation model training, we perform the same experiment but in a multi-task, multi-dataset setting. The joint task comprises band gap, Fermi energy ($\zeta$), and formation energy ($E_\mathrm{form}$) regression along with stability classification for the Materials Project, in addition to formation energy prediction for the Carolina Materials Database. The results shown in Table \ref{tab:wholehog} provide a different perspective on pretraining from the simple task: when trained jointly against multiple types of targets, the general performance for the pretrained model is significantly better than training from scratch, with lower errors for three of the five metrics, with the two remaining being relatively comparable in magnitude. While a more in-depth investigation into the learning dynamics is likely needed, this result does indicate that there is merit in pretraining for foundation models in materials science.

\section{Conclusions and Future Work}

In this work, we have provided a brief introduction into the capabilities of Open MatSci ML Toolkit, focusing on areas relevant to foundation models for materials science. Throughout, we have demonstrated a new pretraining pipeline at scale, using up to 32 compute nodes for fast experimentation turnover. As part of our pretraining experiments, we reported significant training instabilities correlating with large batch training, which we discuss in the context of work by \cite{molybogTheoryAdamInstability2023} underscoring shortcomings in the Adam algorithm. Our results indicate that, when considering pretraining, additional attention may be required balancing training throughput and model convergence when using Adam-like optimizers, and more work is needed into finding optimal learning schedules.

With the pretrained model, we briefly discuss a proof-of-concept application for semantic dataset design. Using a UMAP visualization based on embedded structures across datasets supported in Open MatSci ML Toolkit, we highlight gaps in data space that could be filled to improve the uniformity of datasets---in both chemical and structural spaces---used to train foundation models. Finally, we investigate whether there are tangible benefits with pretraining for downstream tasks relevant to materials science. We find that in the simpler case of single target regression, pretraining can improve early convergence rates but does not result in a more general model. However, for more complex instances such as multi-task, multi-dataset cases relevant to foundation model training, pretraining results in a far more generalized model across targets and datasets than training from random initialization. We hope our early contributions should hopefully guide future experimentation and development of foundation models for materials science, both in terms of data composition and training strategies.

\newpage

\section*{Notices \& disclaimers}

{\footnotesize Performance varies by use, configuration and other factors. Learn more on the Performance Index site. Performance results are based on testing as of dates shown in configurations and may not reflect all publicly available updates. See backup for configuration details.  No product or component can be absolutely secure.  Your costs and results may vary.  Intel technologies may require enabled hardware, software or service activation. \textcopyright~Intel Corporation. Intel, the Intel logo, and other Intel marks are trademarks of Intel Corporation or its subsidiaries. Other names and brands may be claimed as the property of others.}

\bibliographystyle{ACM-Reference-Format}
\bibliography{references}

\appendix

\section{Model descriptions \label{sec:models}}

The encoder backbone used throughout this work is the E(n)-GNN architecture \cite{satorrasEquivariantGraphNeural2022,batznerEquivariantGraphNeural2022a}, which is designed to model equivariance in $n$-dimensions; for chemical applications, equivariance in three-dimensions allows parameterizing functions that commute in euclidean space. Within this framework, model outputs are equivalent for a given set of translation, rotation, and reflection operations, regardless of their sequence ordering. We defer interested readers to \citet{satorrasEquivariantGraphNeural2022}, and only briefly explain the architecture at a high level. Given a graph $\mathcal{G}$ comprising nodes (atoms) $\mathcal{V}$ and edges (bonds) $\mathcal{E}$, the Equivariant Graph Convolutional Layer \cite{satorrasEquivariantGraphNeural2022} transforms node $\mathbf{h}$ and coordinate $\mathbf{x}$ embeddings through message function $\mathbf{m}$ between two nodes $i,j$:

\begin{equation}
    \mathbf{m}_{ij} = \phi_e (\mathbf{h}_i, \mathbf{h}_j, || x_i - x_j ||^2, a_{ij})
\end{equation}

The coordinate embeddings for node $i$ in a subsequent layer ($l+1$) are then given by:

\begin{equation}
    \mathbf{x}^{l+1}_i = x^l_i + C\sum_{j\neq i}(x^l_i - x^l_j) \phi_x (\mathbf{m}_ij)
\end{equation}

where $C$ normalizes the sum aggregation by the dimensionality of the embeddings. E(n)-GNN differentiates from conventional graph convolution by accounting for $x^l_{ij}$ in node and coordinate embeddings given by the last two equations---by taking capturing absolute and relative distances between nodes, thereby preserving translation and rotational equivariance. For our work, the input features comprise atom embeddings from learnable embedding tables, and a vector of $x,y,z$ coordinates, which are passed sequentially through three E(n)-GNN layers, yielding equivariant representations based off three-hop neighborhoods of nodes. Each step combines modeling capabilities from three fully-connected layers with respective nonlinearities to transform each type of embedding: globally, we use \verb|SiLU| activations; for node and message layers, a width of 256 neurons; for positional embeddings, widths of 64. Residual connections are included between graph layers. Finally, at a graph-level, we chose to perform size-extensive modeling, whereby the graph embedding is obtained by summation over the individual node embeddings.

The system-level embeddings are subsequently passed into blocks of multilayer perceptrons (MLP) to transform graph (or point cloud) embeddings into each individual target within a prescribed task. Their design was to strike a balance in the capacity of each output head as to be expressive enough to map onto their targets, while constrained to not ignore the embedding completely. Each output head comprises a sequence of MLP blocks with residual connection, which in turn correspond to MLP\textrightarrow non-linearity\textrightarrow normalization\textrightarrow dropout. The output of the block is subsequently added with the inputs. In the present work, MLPs typically have a hidden dimension of 256, using \verb|SELU| activation \cite{klambauerSelfNormalizingNeuralNetworks2017} with \verb|RMSNorm| \cite{zhangRootMeanSquare2019}, and a dropout probability of 0.2. We did not perform extensive hyperparameter tuning throughout this work; the activation was chosen for slightly better performance over \verb|SiLU|, and \verb|RMSNorm| provided more reliable behavior when operating in the multi-task, multi-data setting where irregular batches may cause issues with traditional \verb|BatchNorm|. For pretraining and fine-tuning in the single task setting, three output blocks were used. For the multi-task, multi-dataset case, each output head comprised six output blocks.

% Description of output heads, and configurations used for pretraining versus task-tuning.

\section{Training curves}

\begin{figure}
    \centering
    \includegraphics[width=0.9\linewidth]{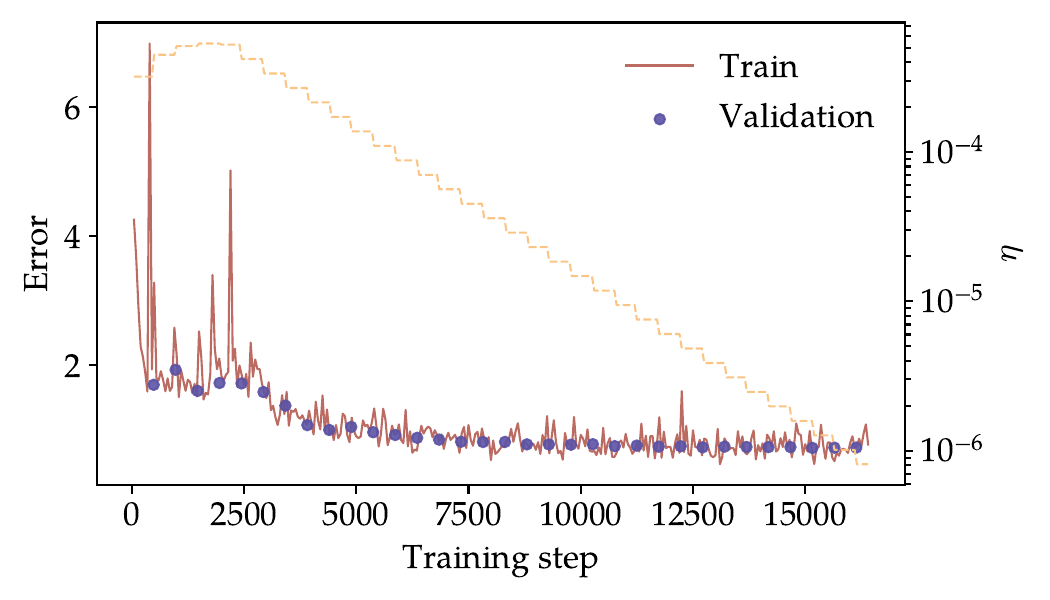}
    \caption{Learning curves for the pretrained E(n)-GNN model used for subsequent fine-tuning experiments, measured by the multiclass cross-entropy error. The yellow dashed line corresponds to the monitored learning rate $\eta$ plot on a logarithmic scale, based on $\eta_\mathrm{base}=10^{-5}$ and $N=512$. One epoch comprises approximately 500 training steps.}
    \label{fig:finalsymmetry}
\end{figure}

\begin{figure*}[ht]
    \centering
    \includegraphics[width=\linewidth]{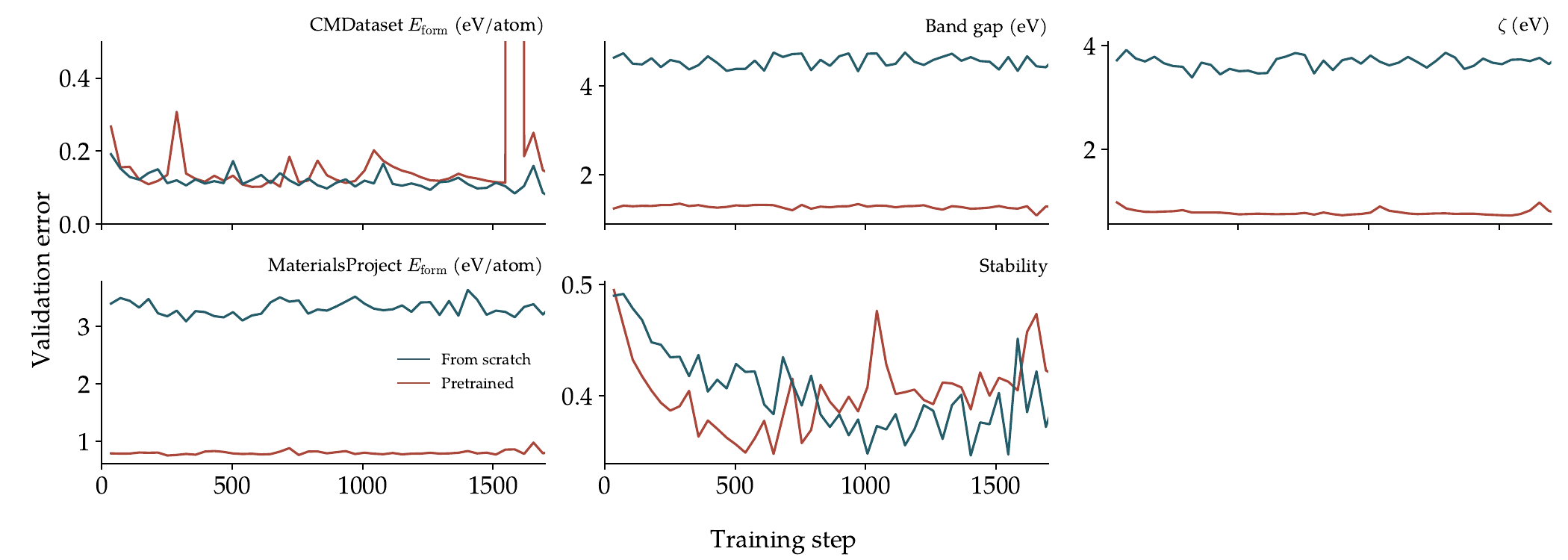}
    \caption{Validation (lower is better) curves for each metric for the multi-task, multi-dataset experiments comparing against models pretrained and from random initialization. For regression targets, units are provided, while ``stability'' is the binary cross-entropy error. For the formation energy, $E_\mathrm{form}$, we differentiate between data from the Carolina Materials Database (\texttt{CMDataset}) and the Materials Project. The $E_\mathrm{form}$ panel for Carolina Materials Database truncates the ordinate, where as the loss spikes to abnormal levels before recovering.}
    \label{fig:wholehog-curves}
\end{figure*}

Figure \ref{fig:finalsymmetry} shows the training curve for the final pretrained E(n)-GNN model used for downstream tasks. In the early stages of training, we observe the training spikes described in Section \ref{sec:pretraining}, which we attributed to the combination of large effective batch sizes, and instabilities owing to the Adam optimizer. The learning rate curve shows the initial linear ramp up to $\eta_\mathrm{base} N$ over five epochs, followed by exponential decay. As the learning rate is decreased, the optimizer stabilizes albeit learning gradually plateaus.

Figure \ref{fig:wholehog-curves} presents the validation error across training, where the final validation errors are shown in Table \ref{tab:wholehog}. We see that for the three cases where pretraining significantly outperforms training from scratch, the model generally struggles to learn throughout training but the pretraining model contains sufficient inductive bias to provide a better baseline. 

\end{document}